\documentclass[nonacm=true,authordraft=false,screen=true,format=sigconf]{acmart}
\usepackage{amsmath}
\usepackage{mathtools}
\usepackage{tikz}
\usepackage{pgf}
\usepackage{pgf-umlsd}
\usepackage{cleveref}
\usepackage{xifthen}
\usepackage{verbatim}

\newboolean{draftMode}
\setboolean{draftMode}{false} 
\ifthenelse{\boolean{draftMode}}{%
  \renewenvironment{sequencediagram}{\fbox{Draft}\comment}{\endcomment}}

\DeclarePairedDelimiter\set\{\}

\newcommand{\RNG}[0]{\text{\it RNG\/}}
\newcommand{\CSRNG}[0]{\text{\it CSRNG\/}}
\newcommand{\bbkdf}[0]{\text{BBKDF}}
\newcommand{\norm}[0]{\text{norm}}
\newcommand{\centralize}[0]{\text{centralize}}
\newcommand{\vectorize}[0]{\text{vectorize}}
\newcommand{\mac}[0]{\text{MAC}}
\newcommand{\thr}[0]{\text{Thr}}

\newcommand{\vecV}[0]{\boldsymbol{V}}
\newcommand{\vecN}[0]{\boldsymbol{N}}

\newcommand{\vecC}[0]{\boldsymbol{C}}
\newcommand{\vecB}[0]{\boldsymbol{B}}
\newcommand{\vecX}[0]{\boldsymbol{X}}
\newcommand{\vecY}[0]{\boldsymbol{Y}}
\newcommand{\vecA}[0]{\boldsymbol{A}}

\newcommand{\concat}[0]{\mathrel{\|}}
\newcommand{\key}{\text{key}}

\ccsdesc[500]{Security and privacy~Key management}
\ccsdesc[500]{Security and privacy~Biometrics}
\ccsdesc[500]{Security and privacy~Privacy-preserving protocols}

\keywords{%
  Authenticated Key Exchange, %
  Biometric Authentication, %
  Cryptography%
}
\copyrightyear{2024}

\makeatletter
\@ifundefined{SetWatermarkText}{}{
  \SetWatermarkText{}}
\makeatother

\title{oBAKE: an Online Biometric-Authenticated Key Exchange Protocol}
\author{Haochen M. Kotoi{-}Xie}
\affiliation{%
  \institution{AnchorZ Inc.}
  \streetaddress{Taitoku Asakusabashi 3-22-9, Daiichi{-}Nomura Bldg. 2F}
  \city{Tokyo}
  \country{Japan}
  \postcode{111-0053}}
\email{kotoi@anchorz.co.jp}
\author{Takumi Moriyama}
\affiliation{%
  \institution{AnchorZ Inc.}
  \streetaddress{Taitoku Asakusabashi 3-22-9, Daiichi{-}Nomura Bldg. 2F}
  \city{Tokyo}
  \country{Japan}
  \postcode{111-0053}}
\email{moriyama@anchorz.co.jp}

\titlenote{%
  This article is a technical note %
  derived from AnchorZ Inc.'s development work of the %
  DZ Intelligent Access Platform.
}

\begin{abstract}
  In this writing, we introduce %
  a novel biometric-authenticated key exchange protocol %
  that allows secure and privacy-preserving key establishment %
  between a stateless biometric sensing system and %
  a ``smart'' user token that possesses biometric templates %
  of the user.

  The protocol yields a shared secret incorporating random nonce from both parties %
  when they positively authenticate each other.
  Mutual positive authentication here is defined as when %
  the feature vector calculated from the sensor data %
  captured by the biometric sensing system %
  only differs from the feature vector stored as the biometric template %
  within the user token %
  by less than a predefined threshold.
  The parties exchange only randomized data and cryptographically derived verifiers; %
  no significant information regarding the vectors is ever exchanged.
  The protocol essentially utilizes the BBKDF~\cite{Seo2018} scheme %
  for feature vector matching, and as a result, the threshold is compared %
  per component of the two vectors to be matched.
  This fact makes it straightforward to employ multiple biometric modalities.

  The protocol also allows online authentication where the biometric sensing system %
  can potentially send multiple queries derived from different sensor data samples, %
  in one or more rounds.
  The protocol is designed in such a way that the user token can very efficiently %
  answer a multitude of such queries.
  This makes the protocol especially suitable for interactive systems %
  while posing a minimal computational burden on the user token.
  The biometric sensing system can be made stateless, i.e. %
  user registration in advance is not required.

  Furthermore, the protocol is bidirectionally privacy-preserving in the sense that %
  unless mutual authentication is achieved first, %
  neither the biometric sensing system, %
  nor the user token %
  can gain useful information, %
  respectively regarding the biometric template, %
  or sensor-data-derived feature vectors. %
\end{abstract}

\begin{document}
\maketitle

\begin{figure}[h]
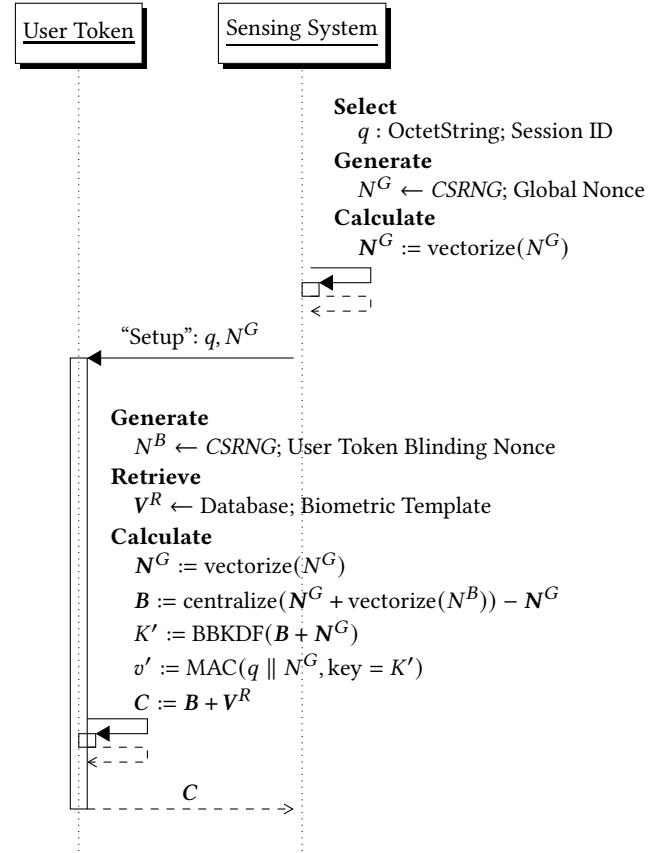

  \centering
\begin{sequencediagram}
  \newinst{AUT}{\shortstack{User Token}}
  \newinst[1]{SUA}{\shortstack{Sensing System}}
  \postlevel
  \postlevel
  \postlevel
  \begin{call}%
    {SUA}{\shortstack[l]{%
        {\bf Select}\\
        \quad $q: \text{OctetString}$; Session ID\\
        {\bf Generate}\\
        \quad $N^G \leftarrow \CSRNG$; Global Nonce\\
        {\bf Calculate}\\
        \quad $\vecN^G := \text{vectorize}(N^G)$
      }}%
    {SUA}{}
  \end{call}
  \begin{call}{SUA}{``Setup'': $q, N^G$}{AUT}{$\vecC$}
    \postlevel
    \postlevel
    \postlevel
    \postlevel
    \postlevel
    \postlevel
    \postlevel
    \begin{call}%
      {AUT}{\shortstack[l]{%
          {\bf Generate}\\
          \quad $N^B \leftarrow \CSRNG$; User Token Blinding Nonce\\
          {\bf Retrieve}\\
          \quad $\vecV^R  \leftarrow \text{Database}$; Biometric Template\\
          {\bf Calculate}\\
          \quad $\vecN^G := \text{vectorize}(N^G)$\\
          \quad $\vecB := \centralize(\vecN^G + \text{vectorize}(N^B)) - \vecN^G$\\
          \quad $K' := \bbkdf(\vecB+\vecN^G) $\\
          \quad $v' := \mac(q \concat N^G,\key=K')$\\
          \quad $\vecC := \vecB + \vecV^R$
        }}%
      {AUT}{}
    \end{call}
  \end{call}
\end{sequencediagram}
\caption{oBAKE Protocol (Phase: Session Initialization)}
\label{fig:seq-init}
\end{figure}

\begin{figure*}[p]
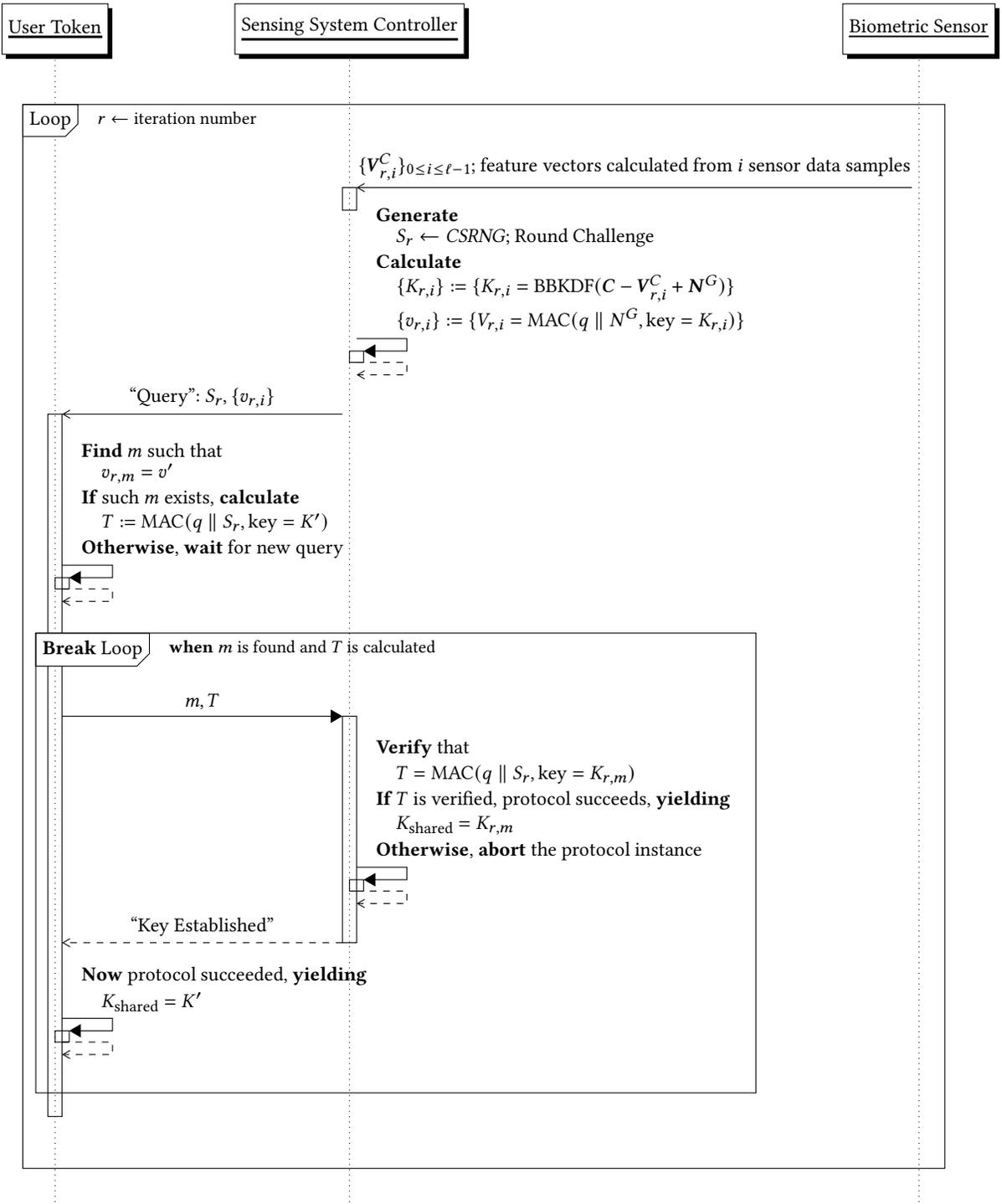

\begin{minipage}{\textwidth}
  \centering
\begin{sequencediagram}
  \newinst{AUT}{User Token}
  \newinst[2]{SUA}{Sensing System Controller}
  \newinst[6]{Sensor}{Biometric Sensor}
  \begin{sdblock}{Loop}{$r \leftarrow \text{iteration number}$}
    \begin{messcall}{Sensor}{%
        $ \set{ \vecV^C_{r,i} }_{0\leq i\leq \ell-1} $; feature vectors calculated from $i$ sensor data samples
      }{SUA}
    \end{messcall}
    \postlevel
    \postlevel
  \begin{call}%
    {SUA}{\shortstack[l]{%
        {\bf Generate}\\
        \quad $S_r \leftarrow \CSRNG$; Round Challenge\\
        {\bf Calculate}\\
        \quad $\set{K_{r,i}} := \set{ K_{r,i} = \bbkdf(\vecC-\vecV^C_{r,i}+\vecN^G)} $\\
        \quad $\set{v_{r,i}} := \set{ V_{r,i} = \mac(q \concat N^G, \key=K_{r,i})}$
      }}%
    {SUA}{}
  \end{call}
  \begin{messcall}{SUA}{``Query'': $S_r, \set{v_{r,i}}$}{AUT}
    \postlevel
    \postlevel
    \postlevel
    \begin{call}%
      {AUT}{\shortstack[l]{%
          {\bf Find} $m$ such that\\
          \quad $v_{r,m} = v'$\\
          {\bf If} such $m$ exists, {\bf calculate}\\
          \quad $T := \mac(q \concat S_r, \key=K')$\\
          {\bf Otherwise}, {\bf wait} for new query
        }}%
      {AUT}{}
    \end{call}
    \begin{sdblock}{{\bf Break} Loop}{{\bf when} $m$ is found and $T$ is calculated}
      \begin{call}{AUT}{$m,T$}{SUA}{ ``Key Established'' }
    \postlevel
    \postlevel
    \postlevel
        \begin{call}%
          {SUA}{\shortstack[l]{%
              {\bf Verify} that\\
              \quad $T = \mac(q \concat S_r,\key=K_{r,m})$\\
              {\bf If} $T$ is verified, protocol succeeds, {\bf yielding}\\
              \quad $K_\text{shared} = K_{r,m}$\\
              {\bf Otherwise}, {\bf abort} the protocol instance
            }}%
          {SUA}{}
        \end{call}
      \end{call}
    \postlevel
    \begin{call}%
      {AUT}{\shortstack[l]{%
          {\bf Now} protocol succeeded, {\bf yielding}\\
          \quad $K_\text{shared} = K'$
        }}%
      {AUT}{}
    \end{call}
    \end{sdblock}
  \end{messcall}
  \end{sdblock}
\end{sequencediagram}
\end{minipage}
\caption{oBAKE Protocol (Phase: Online Authenticated Key Exchange)}
\label{fig:seq-main}
\end{figure*}

\section{Overview}

When a user has his/her biometric templates stored in a token which can be %
carried around, it is desirable at times for digital devices other than the %
biometric-template-carrying user token to authenticate the user by matching %
newly captured biometric sensing data against the template stored in the user token %
\emph{on the spot} (i.e., without the need of prior setup per user--digital device pair), %
and thereafter establish a secure communication channel backed by the biometric authentication.
Doing so enables use cases where an inserted bank card and entering of PIN %
be replaced with a smartphone sitting in the user's pocket and facial recognition %
when performing cash withdraw at an ATM, %
without the need to compromise the user's privacy by requiring the user's %
biometric template to be uploaded to the operator of the ATM.

In this writing, we describe a protocol that allows such use cases while maintaining %
security, availability, and privacy-preservation of the %
combined authentication--key exchange process.

\section{Protocol Description}

\Cref{fig:seq-init} and~\Cref{fig:seq-main} gives a brief description of the oBAKE, %
Online Biometric-Authenticated Key Exchange, protocol.
Please note that we use the notation $X \leftarrow \RNG$ to mean %
generating a random byte string and name it $X$ from a random number generator.
$X \leftarrow \CSRNG$ is a variant of the same notation where %
the random number generator is required to be cryptographically secure.

The protocol is separated into two phases: %
(1)~Session Initialization, and %
(2)~Online Authenticated Key Exchange.

\paragraph{Phase: Session Initialization}
In the first phase, the \emph{biometric sensing system} initialize the protocol by %
deciding $q$ a session id, and $N^G$ a global nonce, %
that are going to be used throughout the lifetime of the protocol.
$q$ and $N^G$ is then sent to the \emph{user token} to mark the %
start of the protocol.

An $\vecN^G$ is derived from $N^G$ to make it arithmetically compatible with %
feature vectors to be matched for the purpose of authentication.
It is assumed that a $\text{vectorize}$ function is chosen as part of the protocol parameters %
which injectively transform a random number %
to a vector that is arithmetically compatible to feature vectors.

After receiving $q$ and $N^G$, the user token chooses $N^B$ a blinding nonce %
and use the same $\text{vectorize}$ function to turn it into $\vecB$ the blinding vector, %
except it is not exactly how we calculate $\vecB$.
We actually offset $\vectorize(N^B)$ use a technique which we developed in association to %
utilize BBKDF in production, which we dub as ``centralization''.
Applying this technique to centralize for $\vecN^G$, %
the actual blinding vector is calculated as %
$\vecB = \centralize(\vecN^G + \vectorize(N^B)) - \vecN^G$, as shown in~\Cref{fig:seq-init}.
See~\Cref{ssec:centralization} for more details about the centralization operation.

After $\vecB$ is derived, user token pre-compute %
the presumptive shared key $K'$ and verifier $v'$ %
as $K' = \bbkdf(\vecB + \vecN^G)$ %
and $v' = \mac(q\concat N^G, key = K') $, respectively.

During the initialization phase, user token also retrieve the \emph{biometric template} $\vecV^R$ %
from its internal database, %
and calculate the blinded template vector $\vecC = \vecB + \vecV^R$, %
which is sent to the sensing system.

\paragraph{Phase: Online Authenticated Key Exchange}
In the second phase, the sensing system captures biometric sensing data regarding %
the person who operates it, and calculate their respective feature vectors. %
Then, from those vectors, derives verifiers used to query the user token.
These queries can happen in many rounds, as the sensing system may not obtain %
optimal biometric sensing data at the first try; %
and in each round the sensing system may send multiple queries, %
to allow optimization of count of round-trip communication.

We denote each verifier corresponding to every sensor data sample as $\vecV^C_{r,i}$, %
where the superscript $^C$ means ``captured'' %
(in comparison to the superscript $^R$ in $\vecV^R$, which means ``registered''), %
$r$ signifies the round index, %
and $i$ differentiate multiple queries within a single round.

For each $\vecV^C_{r,i}$ we derive %
$ K_{r, i} = \bbkdf(\vecC - \vecV^C_{r,i} + \vecN^G)$,
the presumptive shared key,
and a verifier $v_{i,r} = \mac(q\concat N^G, key = K_{r, i})$, %
under the hope where $\norm( \vecV^C_{r,i} - \vecV^R ) < \thr$, %
that is the captured-data-derived feature vector %
is close enough to the user's registered feature vector: %
only differ by no more than a predefined threshold $\thr$.
If those two vectors are indeed close enough, we will have %
$v_{r,i} = v'$, as $ K_{r,i} = K' $ due to the property of %
$\text{BBKDF}$ and $\text{centralize}$ as discussed in~\Cref{ssec:centralization},
and since %
\[
  \begin{aligned}
    \vecC - & \vecV^C_{r,i} + \vecN^G \\
    &= \vecB + \vecV^R - \vecV^C_{r,i} + \vecN^G \\
    &=  \vecV^R - \vecV^C_{r,i} + \centralize(\vecN^G + \vectorize(N^B)) - \vecN^G + \vecN^G \\
    &=  \vecV^R - \vecV^C_{r,i} + \centralize(\vecN^G + \vectorize(N^B))
  \end{aligned}
\] %
also that $K' = \bbkdf(\vecB + \vecN^G)$, %
and that $\vecB + \vecN^G = \centralize(\vecN^G + \vectorize(N^B))$.

After calculating the many $v_{r,i}$ for the certain round, %
the sensing system sends them to the user token to check for potential matches.
The sensing system also include a freshly generated random number $S_r$ as %
the \emph{round challenge}, used to authenticate the user token.

Upon receiving $\set{v_{r,i}}_i$, the user token looks for a match %
against it's pre-computed $v'$.
If one is found, which we denote as $v_{r,m}$, %
the user token can now consider the sensing system as authenticated, %
as the latter must have obtained a $\vecV^C_{r,i}$ that is close enough to $\vecV^R$, %
in other words, the condition for authentication has achieved %
and demonstrated to the user token by the fact that sensing system %
is able to derive $v_{r,m} = v'$.

Next, the user token informs the sensing system that $v_{r,m}$ is a match %
by sending the index $m$, as well as a verifier $T$ to demonstrate its knowledge about %
$ K' $ (which is supposedly equal to $ K_{r,i} $).
$T$ is calculated as $ \mac(q\concat S_r, key = K') $ %
and upon receiving it, the sensing system can verify it by %
comparing it with the expected value producible by $ \mac(q\concat S_r, key = K_{r,m}) $.
After $T$ is verified, it is conventional for the sensing system to %
signal the success of the key exchange process; %
though this is not always necessary if the shared key is used in %
subsequent protocols and therefore the success of the oBAKE protocol %
can be implicitly inferred.

If everything goes according to plan and a matching verifier $v_{r,m}$ is found, %
both the user token and the sensing system should end up with a shared key %
$K_\text{shared} = K' = K_{r,m}$.

If there exists no matching verifier for a certain round, %
the user token simply does nothing while awaiting for the next round.

It is worth noting that after the session initialization phase, %
the user token only perform a single byte string comparison in respond to %
each query, except the final one that results in positive authentication.
This fact makes this protocol especially applicable to %
resource constrained user tokens.

Lastly, if $T$ does not verify, the user token must be %
either malfunctioning or malicious.
There for the sensing system must abort the protocol and take precautions %
when potentially connecting to the same user token.

\subsection{Centralization for BBKDF}\label{ssec:centralization}
BBKDF as described in~\cite{Seo2018} essentially divide the vector space of its input %
into ``cubic-cell''s, where each cell's edge has the length equals to 2 times %
the corresponding component of the matching threshold.
BBKDF then guarantees that all vectors sitting in the same cell yield the same key, %
while it is impossible to tell BBKDF results apart from randomly generated byte string %
when the inputs sit in different cells.
The \emph{centralization} technique we developed is to solve the problem where %
when using BBKDF directly, i.e. performing matching of %
two feature vectors $\vecX$ and $\vecY$ by %
calculating their respective derived keys using BBKDF then compare the keys, %
it is possible that the two vectors are close enough w.r.t. the threshold %
but still yield different key, for example if at a particular dimension
(say the $i$-th dimension), %
$\vecX_i = 7$ and $\vecY_i = 9$ while the cell partitioning phenomenon effectively yields %
cells $\{ .., [1, 9), [9, 17), .. \}$
(assuming a threshold for that dimension being 4). %
To solve this problem, we can introduce an auxiliary vector $\vecV$ so that %
$\vecV + \vecX$ will have every dimension at the center of a cell.
This then guarantees $\bbkdf(\vecX + \vecV) = \bbkdf(\vecY + \vecV)$ to always be true.
One way to find this auxiliary $\vecV$ for an arbitrary $\vecX$ %
is to find the cell center of $\vecX + \vecV'$ and subtract $\vecX$ from it.
Here, $\vecV'$ can be any vector given that the partitioning phenomenon is uniform %
across the scalar space, which is always the case in~\cite{Seo2018}.
We denote this calculation as $\centralize(\vecX + \vecV') - \vecV'$, %
where $\text{centralize}$ is the operation that calculates %
the center of the cell its input occupies.
With this notation in hand, we can also see that given any $\vecA$, $\vecB$ where %
$\norm(\vecA - \vecB)$ is less that the predefined threshold, %
for any $\vecC$ it holds true that $\bbkdf(\vecA - \vecB + \centralize(\vecC))$, %
as $\vecA - \vecB + \centralize(\vecC)$ will always occupy the same cell as %
$\centralize(\vecC)$.

\section{Conclusion}

In this writing, we introduced and described a biometric-authenticated key exchange protocol %
that is online and does not require prior setup per user--digital device pair.

\bibliographystyle{ACM-Reference-Format}
\bibliography{bibiography.bib}

\end{document}